\documentclass[aps,prd,preprint,groupedaddress,showpacs]{revtex4-1}
\usepackage{amsmath}
\usepackage{amsfonts}
\usepackage{graphicx}

\newcommand{\B}{\mathbf} 

\newcommand{\G}{G}
\newcommand{\lam}{\lambda}
\newcommand{\coeff}{\xi}
\newcommand{\q}{q}
\newcommand{\Q}{Q_L}
\newcommand{\flux}{\varphi}

\newcommand{\der}[2]{\frac{\partial#1}{\partial#2}}

\newcommand{\tF}{\tilde{F}}
\newcommand{\tS}{\tilde{\Sigma}}
\newcommand{\non}{\nonumber\\}
\begin{document}
\title{Dissipative String Fluids}
\author{Daniel Schubring}
\email{schub071@d.umn.edu}

\affiliation{Department of Physics, University of Minnesota, Duluth, Minnesota, 55812}
\begin{abstract}
Systems of interacting networks of strings such as cosmic strings or quantum vortices can be approximated in a certain regime as an anisotropic fluid with an equation of state depending on a conserved flux. The equations for ideal magnetohydrodynamics are shown to be another example of a fluid of this type. Previous work on these fluids is now extended to include dissipative effects. The new dissipative terms are discussed in terms of both standard resistive magnetohydrodynamics and small-scale structure formation in networks of cosmic strings. The requirement of frame invariance is shown to restrict the form of higher order corrections to heat flow in the anisotropic direction.
\end{abstract}
\maketitle
\section{Introduction}

Networks of one-dimensional strings appear in a variety of contexts. In particular, networks of quantized vortex lines appear in turbulent quantum fluids, and networks of cosmic strings may have formed in a symmetry breaking phase transition in the early universe. These networks have been extensively studied using numerical models which track the motion of individual strings in the network, as in for instance the vortex-filament model of Schwartz \cite{Schwartz} or the Smith-Vilenkin model for cosmic strings \cite{smithVilenkin}. For many purposes it may be useful to instead consider a `macroscopic' perspective in which individual strings are coarse-grained in a fluid approximation. In the context of quantum turbulence, such an approximation underlies the Hall-Vinen-Bekharevich-Khalatnikhov equations \cite{hvbk} which describe the net vorticity of the network as a continuous field interacting with the usual two-fluid model of a superfluid. On the other hand, in the cosmic string context the dynamics of the strings themselves are often considered independently from any interaction with external fields. Coarse-graining such a network leads to an independent `string fluid' which may exhibit interesting properties distinct from any additional interactions with other fluids.

The individual strings in the network carry a conserved flux. For instance the vortex lines in a superfluid carry quantized angular momentum and the topological defects in the Abelian-Higgs model carry magnetic flux. In the coarse-grained fluid the conservation of flux is manifested as the conservation of an antisymmetric tensor $ F $:
\begin{align}
\nabla_\mu F^{\mu\nu} = 0.\label{dF}
\end{align}
In a fluid of strings carrying magnetic flux, $ F $ is just the dual of the electromagnetic field tensor, and the vanishing of its divergence is just a statement of the homogeneous Maxwell equations. But in fact for any fluid of directed strings there is a conservation law for a tensor $ F $ which describes the topological flux of the strings \cite{Fluid}. It is tempting at this point to point out the similarity to magnetohydrodynamics (MHD) which is another example of a fluid with a conserved magnetic flux. A connection between Nambu-Goto strings and MHD has in fact been previously noticed by Olesen \cite{olesen}. In Sec. \ref{mhd} we will show through quite different methods that ideal MHD is a particular case of what we call a `perfect string fluid'. Formally, a coarse-grained network of strings has many similarities with a plasma, but there are differences in the equation of state of the fluid at equilibrium.

Some readers may here question the idea of an equilibrium for cosmic string networks at all. Through reconnection events the small-scale structure on long strings tends to lead to the production of small loops. It was realized early on from numerical simulations that the reverse process whereby small loops attach to long strings is much less effective for densities below a critical density.\cite{smithVilenkin}\cite{sakelVilenkin} Given a minimum energy cutoff beyond which small loops are restricted from fragmenting, most of the energy will flow into loops of energy comparable to the cutoff size. So any equilibrium properties will be cutoff dependent, and thus artificial in a sense. Of course the idea of separating the string dynamics from all other interactions is artificial as well, and loops near the cutoff may leave the system through various decay channels.

But what the same numerical simulations do show is that very different initial conditions will lead to the same cutoff-dependent equilibrium state, which depends on the energy density as well any net flux of the strings through the system space. And the statistics of the equilibrium states in the numerical simulations agree with analytical calculations by Mitchel and Turok \cite{turok1} which involve notions of temperature and entropy for the string networks. The temperature of the equilibrium states remains near the Hagedorn temperature for a very wide range of densities \cite{turok2}. This may suggest that the decay of small loops and wiggles can be accounted for as the flow of heat from a hot string fluid out of thermal equilibrium with the environment.

In any case, in this paper we will restrict our investigation to the dynamics of an isolated string fluid, and take a macroscopic perspective in which an equation of state is given without reference to an underlying string network. Indeed the example of magnetohydrodynamics shows that what we here call a string fluid may have nothing to do with strings at all on a more microscopic level. The requirements of thermodynamics are then shown to lead to dissipative terms in the fluid equations which correspond to the formation of small-scale structure in an underlying string network.

The paper is organized as follows. In Sec. \ref{secPF} the concept of a perfect string fluid is reviewed. A full treatment emphasizing the variational principle satisfied is found in \cite{perfectFluid}, and the concept has also been studied in the context of blackfolds \cite{blackfold}. The dissipative equations will depend on the equation of state in equilibrium, so two particular cases of a perfect string fluid are discussed. In Sec. \ref{secWiggly} an idealized equation of state for a network of Nambu-Goto strings is reviewed. In Sec. \ref{mhd} it is shown that ideal MHD is another example of a perfect string fluid.

Section \ref{sectTDecomp} begins the discussion of dissipative effects by discussing the ambiguities in choosing the flow velocity and field line direction for a general fluid. Given such a choice, the conserved tensors are broken up into equilibrium and dissipative parts. In Sec. \ref{secEntropy} the entropy current is determined, and the positivity of entropy production is used to find the explicit form of the dissipative terms.

The dissipative parts of the energy-momentum tensor are much the same as for an ordinary fluid, but the dissipative parts of the conserved flux tensor are discussed in \ref{secDiss}. Entropy production due to the curvature of the field lines is discussed in terms of plausible effects in an underlying network of cosmic strings. The nonrelativistic limit of the theory is taken and compared to ordinary resistive magnetohydrodynamics. The dissipative correction to the electric field can be seen as resulting from Ohm's law, but there is an additional term coupling the electric field to temperature gradients.

In Sec. \ref{secEquil} necessary conditions for the fluid to be at equilibrium are derived. As for ordinary fluids, there is a timelike Killing vector proportional to the velocity. In the string fluid there is also an irrotational vector field proportional to the field line direction. In Sec. \ref{sec2nd} an extension to a higher order dissipative theory similar to the Israel-Stewart model \cite{israelStewart} is discussed. The equation describing heat flow along a string is corrected to be hyperbolic, and the speed of second sound is calculated for the idealized cosmic string model discussed in Sec. \ref{secWiggly}.

\section{Perfect String Fluids}\label{secPF}

An ordinary perfect fluid involves one or more conserved currents $n_a^\mu$ (indexed by $a$) which represent extensive quantities such as electric charge, particle number, or entropy. The currents flow in the direction of the timelike velocity $u$ of the fluid,
\begin{align}
n_a^\mu = n_a u^\mu,
\end{align}  and we will here use a $(+,-,-,-)$ signature.

The thermodynamics of the fluid is specified by giving the energy density $\rho$ as a function of the magnitudes $n_a $. Then the chemical potentials $m^a $ are defined as
\begin{align}
m^a \equiv \der{\rho}{n_a},
\end{align}
and the pressure $p$ is defined essentially as a Legendre transform,
\begin{align}
\rho = -p + m^a n_a\\
dp = n_a d m^a\label{dp}
\end{align}

Given these quantities, the energy-momentum tensor is just
\begin{align}
T^{\mu\nu}=(\rho+p)u^\mu u^\nu - p g^{\mu\nu},
\end{align}
and the fluid equations are equivalent to the conservation laws
\begin{align}
\nabla_\mu T^{\mu\nu} = \nabla_\mu n_a^\mu = 0.\label{dTn}
\end{align}

Note that if one pair of density and chemical potential is singled out as the entropy density $s$ and temperature $T$, the remaining conservation laws \eqref{dTn} and the expression for the derivatives of the pressure \eqref{dp} can be used to prove the conservation of $s$,
\begin{align}
u_\mu\nabla_\nu T^{\mu\nu} &= \nabla_\nu (m^a n_a + Ts)u^\nu - u^\nu\nabla_\nu p\non
&= T \nabla_\nu su^\nu.\label{ds}
\end{align}
Similar expressions will be useful in extending to the dissipative case.

A string fluid also involves the conservation of at least one antisymmetric flux tensor $F$,
\begin{align}
\nabla_\mu F^{\mu\nu}= 0.
\end{align}
In the case of a perfect string fluid, $F$ is a simple bivector that can be written as the alternating product of two vectors. Further, the fluid velocity $u$ is in the linear space spanned by these vectors. The velocity $u$ can be used to define a normalized spacelike direction $w$ and a positive magnitude $\flux$,
\begin{align}
\flux\,w^\mu&\equiv F^{\mu\nu}u_\nu  \\
u^\mu u_\mu&= -w^\mu w_\mu = 1\\
u^\mu w_\mu &= 0.
\end{align}

Together, $u$ and $w$ determine the directional part $\Sigma$ of $F$,
\begin{align}
\Sigma^{\mu\nu}&\equiv w^\mu u^\nu - u^\mu w^\nu\\
F^{\mu\nu} &= \flux \Sigma^{\mu\nu}.\label{F0}
\end{align}
It will also be useful to define the projector $h$ onto the space spanned by $u$ and $w$, and its orthogonal complement $\perp$,
\begin{align}
h^{\mu}_{\,\,\,\nu}&=\Sigma^{\mu\rho}\Sigma_{\rho\nu}\non
&= u^\mu u_\nu - w^\mu w_\nu\label{h}\\
\perp^{\mu}_{\,\,\,\nu}&=\tS^{\mu\rho}\tS_{\nu\rho}\non
&= \delta^{\mu}_{\,\,\,\nu} -u^\mu u_\nu + w^\mu w_\nu,\label{perp}
\end{align}
where we are using tildes to denote the Hodge dual,
\begin{align}
\tS_{\mu\nu}\equiv\frac{1}{2}\epsilon_{\mu\nu\rho\sigma}\Sigma^{\rho\sigma}.
\end{align}
The dual $\tF$ of $F$ itself is a two-form that can be integrated to give the net flux carried by the strings across a surface. The magnitude $\flux$ is thus a measure of this flux and it is taken to be a thermodynamic variable on the same footing as the densities $n_a$. The conjugate chemical potential to $\flux$ is denoted by $\mu$,
\begin{align}
\mu \equiv \der{\rho}{\flux}.
\end{align}
And the pressure for a string fluid now involves $\mu\flux$,
\begin{align}
\rho = -p + m^a n_a + \mu\flux.
\end{align}

In an earlier paper \cite{perfectFluid} it was shown that a quite general variational principle leads to an energy-momentum tensor of the form
\begin{align}
T^{\mu\nu} = (\rho+p)u^\mu u^\nu - (\tau+p)w^\mu w^\nu - p g^{\mu\nu}\label{t0},
\end{align}
where $\tau$ is a thermodynamic potential related to the tension of the strings,
\begin{align}
\tau \equiv -p + \mu\flux.
\end{align}
The equations of motion of the perfect string fluid are then equivalent to the conservation of $T^{\mu\nu} $ and all currents and fluxes \eqref{dTn}\eqref{dF}.

\subsection{Wiggly string fluid}\label{secWiggly}

We will now review some particular examples of string fluids. Directly coarse-graining a network of Nambu-Goto strings leads to $T $ and $F$ tensors expressed in terms of correlations between the (non-unit vector) string velocity $U$ and the tangent vector to the string $W$.\cite{Fluid}\cite{wigglyFluid}
\begin{align}
T^{\mu\nu}=\langle U^\mu U^\nu- W^\mu W^\nu\rangle\non
F^{\mu\nu}=\langle W^\mu U^\nu- U^\mu W^\nu\rangle.\label{wigglyTF}
\end{align}
The vectors $U$ and $W$ are properties of the individual strings in the network and the brackets denote an integration over a coarse-graining volume. There are sixteen independent components of these tensors, and so the conservation of the $T$ and $F$ tensors alone does not fully specify the system.

The extra assumption needed was suggested by Vanchurin's kinetic theory of a gas of string segments. \cite{Kinetic} This model suggested that the strings would equilibriate to a state in which there are no correlations between the statistics of right and left movers. The principle that the string fluid should everywhere locally be in an equilibrium of this form allowed for the correlations in \eqref{wigglyTF} to be factored into the average string velocity field $\bar{U}$ and the average tangent vector field $\bar{W}$. 

At this point we will note that for a general string fluid the conservation $\nabla_\mu F^{\mu\nu}=0$ together with the condition that $F$ be a simple bivector implies that spacetime can be foliated by two-dimensional manifolds that are everywhere tangent to the linear subspace defined by the projector $h$ in \eqref{h}.\cite{wigglyFluid} Since $\bar{U}$ and $\bar{W}$ lie in this tangent space, it is tempting to interpret the manifolds as the worldsheets of `macroscopic strings' which point in the direction of the field lines of $\bar{W}$ and propagate with velocity $\bar{U}$.

Ultimately these fields can be expressed in terms of the variables $\flux, u, w$ in the present paper, and the energy-momentum tensor takes the form
\begin{align}
T^{\mu\nu}=\flux(M u^\mu u^\nu - T w^\mu w^\nu),
\end{align}
where the quantities $M$ and $T$ can be respectively interpreted as the mass-per-length and tension of the macroscopic strings.

In fact $M$ and $T$ have exactly the same form as the mass-per-length and tension of a single `wiggly string' which can be described as an ordinary Nambu-Goto string with small-scale perturbations integrated out.\cite{VilenkinWiggly}\cite{CarterWiggly} In the string fluid, the wiggles of the macroscopic strings may also involve disconnected loops smaller than the coarse-graining scale. The coarse-grained wiggles appear in the string fluid as a conserved `wiggle number density' $n$, in terms of which the equation of state can be expressed as
\begin{align}
\rho(n,\flux) = \flux M = \sqrt{(\mu_0 \flux)^2 + n^2},
\end{align}
where $\mu_0$ is the mass per length of a Nambu-Goto string.\cite{perfectFluid}

Given that $n$ describes structure below the macroscopic scale, and that the tendency towards production of small loops should monotonically increase $n$, this strongly suggests that $n$ is proportional to the entropy density $s$:
\begin{align}
\rho = \sqrt{(\mu_0 \flux)^2 + (T_H s)^2}\label{wigglyState},
\end{align}
where $T_H$ is some constant of proportionality. In the limit as $s$ goes to infinity, the temperature $T$ goes to the finite value $T_H$,
\begin{align}
T\equiv \left(\der{\rho}{s}\right)_\flux \rightarrow T_H,
\end{align}
which suggests that we identify $T_H$ as the Hagedorn temperature. For a single wiggly string there is also a corresponding conserved current and equation of state (differing by a factor of $\flux$), and the identification of this current as the entropy has been previously made \cite{CarterWarm}. Even so the entropy is conserved in both the dynamics of wiggly strings and in perfect string fluids. The idea will be extended in this paper by introducing dissipative effects leading to increases in entropy density.

\subsection{Magnetohydrodynamics}\label{mhd}

A relativistic formulation of magnetohydrodynamics is given for instance by Harris \cite{MHDHarris}.. The energy-momentum tensor is simply the sum of a fluid part and an electromagnetic part,
\begin{align}
T^{\mu\nu}&=T_{\text{m}}^{\mu\nu}+T_{\text{EM}}^{\mu\nu}\non
&=(\rho+p)u^\mu u^\nu - p g^{\mu\nu} - \tF^{\mu\rho}\tF^{\nu}_{\,\,\,\rho}+\frac{1}{4}g^{\mu\nu}\tF^{\rho\sigma}\tF_{\rho\sigma}.
\end{align}
Taking the divergence,
\begin{align}
u_\nu\nabla_\mu T^{\mu\nu} = T\nabla_\mu su^\mu  - u_\nu\tF^{\nu}_{\,\,\,\rho}j^\rho = 0, \label{dsMH}
\end{align}
where we have used the homogenous Maxwell equations \eqref{dF} and the expression for divergence of entropy \eqref{ds}, and the current $j$ is defined by the Maxwell equations,
\begin{align}
j^\mu \equiv \nabla_\mu\tF^{\mu\rho}.
\end{align}
The positivity of entropy production,
\begin{align}
\nabla_\mu su^\mu \geq 0,
\end{align} will be satisfied if in fact the current is given by
\begin{align}
j^\rho  = q u^\rho + \sigma \tF^{\mu\rho}u_\mu,
\end{align}
where $\sigma$ is a positive scalar and $q$ can be arbitrary. But in the rest frame of the fluid $\tF^{\mu\nu}u_\mu$ is just the electric field, so this is just a statement of Ohm's law \cite{Eckart}. We will later return to this point, but presently we will consider the isentropic case of ideal magnetohydrodynamics.

For entropy to be conserved in \eqref{dsMH} the electric field must vanish in the rest frame,
\begin{align}
\tF^{\mu\nu}u_\nu = 0.\label{eVanish}
\end{align}
This is just the well-known condition for frozen-in magnetic field lines, but for our purposes it implies that $\tF$ and its dual $F$ are simple bivectors, and that $u$ is in the linear subspace spanned by $F$. So we can define $\flux$ and $w$ as before, noting that they can be interpreted as the magnitude and direction of the magnetic field in the rest frame.

The energy-momentum tensor can be simplified using the expression for the orthogonal projector \eqref{perp},
\begin{align}
T^{\mu\nu}&=(\rho+p)u^\mu u^\nu - p g^{\mu\nu} - \flux^2\perp^{\mu\nu}+\frac{1}{2}g^{\mu\nu}\flux^2\non
&= (\rho + p + \flux^2)u^\mu u^\nu - \flux^2 w^\mu w^\nu -(p+\frac{1}{2}\flux^2)g^{\mu\nu}.
\end{align}
So if the total equation of state is taken as
\begin{align}
\rho_\text{total} = \rho + \frac{1}{2}\flux^2,
\end{align}
then the other thermodynamic quantities are found to be
\begin{align}
\mu &= \flux\label{mhdMu}\\
p_\text{total} &= p+\frac{1}{2}\flux^2\\
\tau+p_\text{total} &= \flux^2,
\end{align}
showing that this is indeed an example of a perfect string fluid.

Note that the form of the energy density is just what we would expect from the variational principle for perfect string fluids \cite{perfectFluid}. There it was shown that the total energy density ends up being the negative of the Lagrangian. And the extra term in the energy density is just the negative of the usual Lagrangian for electromagnetism 
\begin{align}
-\frac{1}{4}\tF^{\rho\sigma}\tF_{\rho\sigma}=-\frac{1}{2}\flux^2.\nonumber
\end{align}
\section{Dissipative String Fluids}
\subsection{Tensor decomposition}\label{sectTDecomp}\begin{flushright}

\end{flushright}
In a more general string fluid the conservation equations for $T$, $F$, and any additional conserved currents $n_a$ still hold, but the tensors are no longer in the equilibrium forms \eqref{F0}\eqref{t0}. Just as for an ordinary dissipative fluid, there is no longer a single preferred fluid velocity $u$. We may take the fluid velocity to be in the direction of the timelike eigenvector of energy-momentum tensor (a choice known as the `Landau-Lifshitz frame' \cite{LL}) or we may choose the velocity to be in the direction of one of the currents (known as the `Eckart frame' \cite{Eckart}) ---the directions no longer coincide in general. In a string fluid we are now faced with the additional problem that the tensor $F$ may no longer be a simple bivector, and so there is ambiguity in how to define $w$.

We may still select $u$ and $w$ as orthonormal vectors in the two-dimensional timelike eigenspace of $F^{\mu\rho}F_{\rho\nu}$. In general the fluid velocity from the Eckart or Landau-Lifshitz frames will not lie in this space so this can define a distinct third possible choice for velocity. As we will see, this frame will have some similarities to the Eckart frame. To distinguish the two cases, the ordinary Eckart frame will be referred to as the `particle frame' and the choice of velocity from this eigenspace as the `string frame'.

There is also the difficulty that none of the frames above satisfy the integrability conditions of the perfect string fluid. We can no longer foliate spacetime by worldsheets everywhere tangent to $u$ and $w$. However the conservation of $F$ does imply that we can define a gauge potential $A$,
\begin{align}
\tF \equiv dA.\nonumber
\end{align}
And by Darboux's theorem $ A $ can be written in terms of four scalar fields $X_1,Y_1,X_2,Y_2$,
\begin{align}
A \equiv X_1 \,dY_1 + X_2 \,dY_2.\nonumber
\end{align}
So then $\tF$ can be decomposed into two simple two-forms with vanishing exterior derivatives
\begin{align}
\tF = dX_1 \wedge dY_1 + dX_2 \wedge dY_2.\nonumber
\end{align}
These two-forms each annihlate a two-dimensional space which does satisfy the integrability condition. So this could be used to define yet another natural choice of $u$ and $w$ which preserves the integrability condition.

We restrict our attention to fluids that are sufficiently close to equilibrium so that the difference between these frames is `small'. We will be more precise on this point later, where frame invariance will be used to restrict higher order dissipative terms in the theory. For now, given a choice of $u$ and $w$, we can define $\rho$, $\flux$, and $n_a$ from the nonequilibrium tensors,
\begin{align}
\rho &\equiv T^{\mu\nu}u_\mu u_\nu\non
\flux &\equiv F^{\mu\nu}u_\mu w_\nu\non
n_a &\equiv n_a^\mu u_\mu.\label{rho}
\end{align}
These values can be used to define the other thermodynamic quantities through the equilibrium equation of state. And so $T$ and $F$ can be decomposed into an equilibrium tensor and a nonequilibrium correction. The nonequilibrium correction may further be decomposed into parts parallel and orthogonal to $u$ and $w$.
\begin{align}
T^{\mu\nu} &=(\rho+p) u^\mu u^\nu - \mu\flux w^\mu w^\nu - p g^{\mu\nu} + 2\q^{(\mu}u^{\nu)} + \pi^{\mu\nu}\label{Tdiss}\\
F^{\mu\nu} &= \flux \Sigma^{\mu\nu} - 2 u^{[\mu}\lam^{\nu]} + 2 w^{[\mu}\nu^{\nu]} + \G^{\mu\nu}\\
n_a^\mu &= n_a u^\mu + N_a w^\mu+\nu_a^\mu,\label{Ndiss}
\end{align}
and $q$ and $\pi$ are further split,
\begin{align}
q^\mu &\equiv \Q w^\mu + q_T^\mu\\
\pi^{\mu\nu}&\equiv -\Pi_L w^{\mu}w^{\nu} + \Pi_T \perp^{\mu\nu} -2w^{(\mu}\pi_L^{\nu)} +\pi_T^{\mu\nu}.\label{pi}
\end{align}
The vectors and tensors $\lambda,\nu,\G,\q_T,\pi_L,\pi_T,\nu_a$ are all fully orthogonal to $u$ and $w$, and $\pi_T$ is defined to be traceless. It should be emphasized that this is simply a decomposition of the tensors, and there is no loss of generality at this point.

If $u$ and $w$ are taken from our preferred frames some of these pieces vanish. In the string Eckart frame, $u$ and $w$ are chosen from an eigenspace so that both vectors $\lam$ and $\nu$ in $F$ vanish. There is still some freedom in our choice of $u$, but there is a unique $u$ such that the longitudinal heat flow $\Q$ vanishes.

In the Landau-Lifshitz frame the vector $\nu$ is nonzero but all heat flow components $q$ vanish. Specifying $w$ through
\begin{align}
\flux w^\mu \equiv F^{\mu\nu}u_\nu,
\end{align}
the vector $\lam$ vanishes as well.

\subsection{Entropy current}\label{secEntropy}
The entropy density $s$ is defined through the equilibrium equation of state, and satisfies the usual thermodynamic identities
\begin{align}
s &= \frac{p}{T} + \frac{1}{T}\rho -\frac{\mu}{T} \flux - \frac{m^a}{T}  n_a\non
ds &= \frac{1}{T} d\rho - \frac{\mu}{T} d\flux  -\frac{m^a}{T} dn_a \label{dsTherm}.
\end{align}
It will be useful to promote the derivatives of the entropy to vectors,
\begin{align}
\beta^\mu &\equiv \frac{1}{T}u^\mu\\
\alpha^\mu &\equiv \frac{\mu}{T}w^\mu.
\end{align}
Then the equilibrium entropy current can be written in terms of the equilibrium tensors $T_0$, $F_0$, $n_{a0}$
\begin{align}
s u^\mu = {p}\beta^\mu + \beta_\nu T_0^{\mu\nu} - \alpha_\nu F_0^{\mu\nu} - \frac{m^a}{T} n_{a0}^\mu\\
d(su^\mu) = \beta_\mu dT_0^{\mu\nu} - \alpha_\nu dF_0^{\mu\nu} - \frac{m^a}{T} dn_{a0}^\mu.\label{s0}
\end{align}

Closely following the approach of Israel and Stewart \cite{israelStewart} we then make the assumption that the derivatives of nonequilibrium entropy current $s^\mu$ satisfy the same relation with the nonequilibrium tensors,
\begin{align}
ds^\mu = \beta_\mu dT^{\mu\nu} - \alpha_\nu dF^{\mu\nu} - \frac{m^a}{T} dn_a^\mu.
\end{align}
The entropy current is taken to be a function of the components of $T,F,n_a$, and we can expand about the equilibrium point $T_0$, $F_0$, $n_{a0}$. To first order,
\begin{align}
s^\mu &= s u^\mu +\beta_\mu (T-T_0)^{\mu\nu}- \alpha_\nu (F-F_0)^{\mu\nu} - \frac{m^a}{T} (n_a-n_{a0})^\mu\non
&=s u^\mu + \frac{1}{T}q^{\mu} - \frac{\mu}{T}\nu^{\mu} - \frac{m^a}{T} \nu_a^\mu.\label{scurrent}
\end{align}
Comparison with \eqref{dsTherm} suggests $q$ is naturally interpreted as a heat vector describing the transport of energy in the rest frame. The currents $\nu$ and $\nu_a$ respectively describe the transport of flux and charge in the rest frame through diffusion.

Expressions for the dissipative quantities appearing in the theory can now be determined by requiring that the entropy production be non-negative
\begin{align}
\nabla_\mu s^\mu \geq 0.\label{secondL}
\end{align} 
The divergence of $s$ can be found through similar manipulations as those leading to the conservation of entropy in the perfect fluid \eqref{ds}. For brevity at this point we will consider a theory with no dependence on conserved currents $n_a$, and choose the Landau-Lifshitz frame so that the heat vector $q$ vanishes. These aspects of the derivation are no different than that for particle fluids (see e.g. \cite{LL}) and can be easily derived for a string fluid in the same way. Beginning with the dissipative energy-momentum tensor \eqref{Tdiss}:
\begin{align}
u_\nu \nabla_\mu T^{\mu\nu} &= \nabla_\mu (\rho+p)u^\mu + \mu\flux w^\mu w^\nu \nabla_\mu u_\nu - u^\mu \nabla_\mu p - \pi^{\mu\nu}\nabla_\mu u_\nu\non
&= T\nabla_\mu su^\mu + \mu \nabla_\mu \flux u^\mu - \mu\flux h^{\mu\nu} \nabla_\mu u_\nu- \pi^{\mu\nu}\nabla_\mu u_\nu\label{deriv}
\end{align}
where $h$ is the projection operator defined in \eqref{h}. If it were still true that $ F = \flux \Sigma $ the middle terms involving $\flux$ would cancel using a relation derived in \cite{perfectFluid}. This would be one way to show entropy is conserved in a perfect string fluid. But now the relation is modified due to dissipative terms in $F$,
\begin{align}
\nabla_\mu\flux u^\mu &= \nabla_\mu( \flux \Sigma^{\mu\lambda}w_\lambda) \non
&=\nabla_\mu(F^{\mu\lambda}w_\lambda - \nu ^\mu)\non
&= F^{\mu\lambda}\nabla_\mu  w_\lambda -\nabla_\mu\nu ^\mu \non
&=\flux \Sigma^{\mu\lambda}\nabla_\mu w_\lambda + (2w^{[\mu}\nu^{\lambda]}+G^{\mu\lambda})\nabla_\mu w_\lambda -\nabla_\mu\nu ^\mu\non
&=\flux h^{\kappa}_{\,\,\,\mu}\nabla_\kappa\Sigma^{\mu\lambda}w_\lambda - \flux w_\lambda h^{\kappa}_{\,\,\,\mu}\nabla_\kappa\Sigma^{\mu\lambda} + \dots.\nonumber
\end{align}
It can be shown (for instance by explicitly writing $\Sigma$ and $h$ in terms of $u$ and $w$) that $h^{\kappa}_{\,\,\,\mu}\nabla_\kappa\Sigma^{\mu\lambda}$ is orthogonal to $w$. So the second term above vanishes, and returning to the derivation \eqref{deriv},
\begin{align}
0& =\nabla_\mu su^\mu + \frac{\mu}{T}[ (2w^{[\mu}\nu^{\lambda]}+G^{\mu\lambda})\nabla_\mu w_\lambda -\nabla_\mu\nu ^\mu ]- \frac{1}{T}\pi^{\mu\nu}\nabla_\mu u_\nu\non
&=\nabla_\mu (su^\mu-\frac{\mu}{T}\nu ^\mu) + \frac{\mu}{T}(2w^{[\mu}\nu^{\lambda]}+G^{\mu\lambda})\nabla_\mu w_\lambda +\nu ^\mu\nabla_\mu\frac{\mu}{T}- \frac{1}{T}\pi^{\mu\nu}\nabla_\mu u_\nu\non
&=\nabla_\mu s ^\mu + \frac{\mu}{T}G^{\mu\lambda}\nabla_\mu w_\lambda +\nu ^\mu(\nabla_\mu\frac{\mu}{T}+\frac{\mu}{T}w^{\lambda}\nabla_\lambda w_\mu)- \frac{1}{T}\pi^{\mu\nu}\nabla_\mu u_\nu.\label{deriv2}
\end{align}
Now the second law \eqref{secondL} will be satisfied if each of the other terms is strictly negative. So we choose $\nu$ and $\G$ to have the form
\begin{align}
\nu^\mu &= \coeff_T \perp^{\mu\rho}(\nabla_\rho \frac{\mu}{T} + \frac{\mu}{T} w^\sigma\nabla_\sigma w_\rho)\label{nu}\\
G^{\mu\nu} &= -\coeff_L \frac{\mu}{T} \perp^{\mu\rho}\perp^{\nu\sigma}\nabla_{[\rho}w_{\sigma]},\label{G}
\end{align}
where the coefficients $\coeff_T,\coeff_L$ are positive scalars. Breaking up the viscous tensor $\pi$ into its parts as in \eqref{pi} we find a series of terms each of which is set to be negative by choosing
\begin{align}
\Pi_L &= -3 \,\zeta_L\, w^\rho w^\sigma \nabla_\rho u_\sigma\\
\Pi_T &= \frac{3}{2} \,\zeta_T \,\perp^{\rho\sigma}\nabla_\rho u_\sigma\\
\pi_L^{\mu}&=2\,\eta_L\,\perp^{\mu\rho}w^\sigma\nabla_{(\rho}u_{\sigma)}\label{piL}\\
\pi_T^{\mu\nu}&=2\,\eta_T\left(\perp^{\mu\rho}\perp^{\nu\sigma}-\frac{1}{2}\perp^{\mu\nu}\perp^{\rho\sigma}\right)\nabla_{(\rho}u_{\sigma)},
\end{align}
with positive coefficients $\zeta_L,\zeta_T,\eta_L,\eta_T $. In principle the physics in the longitudinal direction $w$ may be different from the transverse directions, which is why there are twice as many dissipative coefficients as for an isotropic fluid. The normalization of the coefficients is chosen so that if the physics were isotropic $\zeta_L=\zeta_T$ would be the usual bulk viscosity coefficient and $\eta_L=\eta_T$ the usual shear viscosity coefficient.

Note that the longitudinal viscosity vector $\pi_L$ \eqref{piL}  potentially represents two distinct physical effects. One is due to changes in the transverse velocity along a single macroscopic string or field line. The other effect is due to differences in the longitudinal velocities of nearby strings. Due to the symmetry of the energy-momentum tensor these must be described by the same viscosity coefficient, but if we allow for intrinsic angular momentum these could in principle be different.

For completeness we may also consider a frame in which the heat vector $q$ does not vanish. Following the same line of derivation there would be an extra entropy production term
\[ -q^{\mu}(\nabla_\mu \frac{1}{T} + \frac{1}{T}u^\nu\nabla_\nu u_\mu) \]
in \eqref{deriv2}. The two pieces of $q$ are thus set as
\begin{align}
\Q &= \kappa_L w^\mu (\nabla_\mu T - T u^\nu \nabla_\nu u_\mu)\label{Q}\\
q_T^\nu &= \kappa_T \perp^{\mu\nu} (\nabla_\mu T - T u^\nu \nabla_\nu u_\mu),
\end{align}
where $\kappa_L,\kappa_T$ are the positive heat conductivity coefficients. The apparent difference in sign from the Fourier heat conduction law is just due to the signature of the metric.

\subsection{Dissipation in F}\label{secDiss}

Besides the appearance of an anisotropic direction, the dissipative terms in $T$ are essentially the same as for an ordinary fluid. What may require some interpretation are the dissipative terms \eqref{nu}\eqref{G} in $F$,
\begin{align}
F^{\mu\nu} = 2w^{[\mu}(\flux u + \nu)^{\nu]}  + \G^{\mu\nu}.
\end{align}
The tensor is here written in a form emphasizing the analogy to ordinary particle currents \eqref{Ndiss}.  The velocity $u_E$ in the string Eckart frame where $\nu$ does not appear explicitly in $F$ is clearly given by
\begin{align}
u_E \approx u + \frac{1}{\flux}\nu,\label{eckart}
\end{align}
where this is only an equality to first order in the dissipative fields.
Following a similar line of reasoning to Landau-Lifshitz \cite{LL}, we replace the velocity in the first term of the energy-momentum tensor,
\begin{align}
(\rho+p)u^\mu u^\nu \approx (\rho+p)u_E^\mu u_E^\nu - 2\frac{\rho+p}{\flux}\nu^{(\mu} u_E^{\nu)}.
\end{align}
So the heat vector in the Eckart frame is approximately
\begin{align}
q_E=-\frac{\rho+p}{\flux}\nu.\label{eckartHeat}
\end{align}
Substituting the expression \eqref{nu} for $ \nu $ and ignoring the term due to curvature of $w$,
\begin{align}
q_E=\frac{\rho+p}{\flux}\coeff_T \nabla_\perp \frac{\mu}{T}.
\end{align}
So by the thermodynamic identity
\begin{align}
 Td(\frac{\mu}{T})= -\left(\frac{\rho+p}{\flux T}\right)dT + dp, \label{thermIden}
\end{align}
we can make the identification
\begin{align}
\coeff_T = \left(\frac{\flux T}{\rho+p}\right)^2 \kappa_T.
\end{align}

So $\xi_T$ can be related to heat conductivity ---but this is not the only way to understand $\nu$, and the interpretation of $G$ is still obscure. This may be clarified by taking the nonrelativistic limit:
\begin{align}
\nabla_\mu &= (c^{-1}\partial_t,\nabla_i)\non
u^\mu &\rightarrow (1,c^{-1}\B{v})\\
w^\mu &\rightarrow (c^{-1}\B{v}\cdot\B{w},\B{w}),
\end{align}
where $\B{w}$ is a unit vector. The metric is taken to be the Minkowski metric, so as $c$ goes to infinity the time components of $\perp^{\mu\nu}$ go to zero. Thus the time components of $\nu$ and $G$ vanish, and we will take the spatial components to be of order $c^{-1}$. So in the nonrelativistic limit $\nabla_\mu F^{\mu\nu}=0$ is reduced to the equations
\begin{align}
\nabla\cdot(\flux\B{w})&=0\\
\partial_t (\flux\B{w}) &= -\nabla\times (\flux\B{w}\times \B{v})-\nabla\times (\B{w}\times \B{\nu}) -\nabla_i G^{ij}\label{faraday}.
\end{align}
Using the limit of the spatial part of the projection tensor
\[ \perp^{ij}=-\delta^{ij}+w^i w^j, \]
the dissipative parts are expressed as
\begin{align}
\B{\nu}&=-\coeff_T(\nabla_\perp \frac{\mu}{T}-\frac{\mu}{T}\B{\kappa})\\
G^{ij}&=\coeff_L \frac{\mu}{T} (\nabla^{[i}w^{j]}-w^{[i}\kappa^{j]}),
\end{align}
with the curvature vector
\begin{align}
\B{\kappa}\equiv (\B{w}\cdot\nabla)\B{w}=(\nabla\times\B{w})\times\B{w},
\end{align}
and $\nabla_\perp$ indicates the gradient with the w-component projected out. The curvature also satisfies the identity
\[ \B{w}\times\B{\kappa}=\nabla\times\B{w}- (\B{w}\cdot\nabla\times\B{w})\B{w}, \]
which is used in $\B{w}\times\B{\nu}$ and the dual of $G$,
\begin{align}
\tilde{\B{G}}&=\coeff_L \frac{\mu}{T} (\B{w}\cdot\nabla\times\B{w})\B{w}\non
\B{w}\times\B{\nu}&=-\coeff_T(\B{w}\times\nabla \frac{\mu}{T}-\frac{\mu}{T}(\nabla\times\B{w})_\perp)\label{gtilde}.
\end{align}
We have already discussed how $\coeff_T$ and gradients in $\mu/T$ are related to heat conduction. Now even if the thermodynamic variables are constant notice that $\coeff_T$ and $\coeff_L$ describe the production of entropy due to the curl of the field lines in the transverse and longitudinal directions respectively.

This can be intuitively understood in the wiggly string fluid. A curl that is completely perpendicular to $w$ is found for instance in large loops lying in a plane. The loops tend to contract under tension in the direction of curvature. There is an outflow of heat due to the emission of small loops as the strings contract, so there will still be some net flow of strings $\nu$ even in the rest frame where there is no net flow of energy.

\begin{figure}
\includegraphics[width=0.60\textwidth]{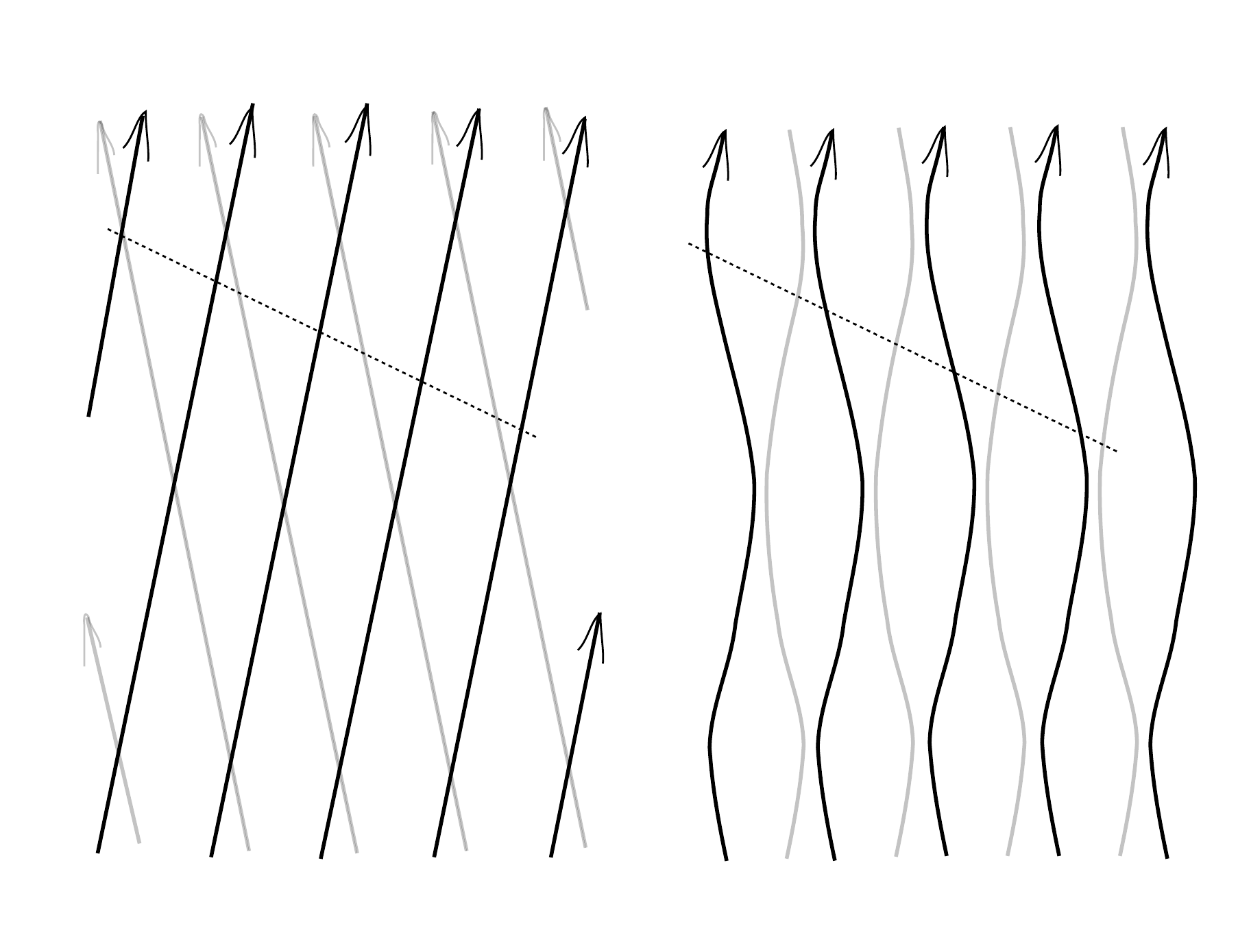}
\caption{Adjacent layers of straight strings diffuse and overlap. Through reconnection, entropy in the form of wiggles is produced. There is also some loss of net flux as indicated by the number of black wiggly strings passing the dotted line.\label{figure}}
\end{figure}

One idealized situation in which only the coefficient $\coeff_L$ applies is when each individual field line of $w $ is an infinite straight line, and all field lines in a given plane perpendicular to some axis are pointing in the same direction. If the direction of the field lines in a plane changes as we move along the axis, the curl of $w$ will point in the direction of $w$ itself. If strings from one plane diffuse to an adjacent layer reconnections will lead to the production of entropy in the form of wiggles and there will be some loss of flux (see Fig.\ref{figure}). This last point is perhaps easiest to understand in the limit of two layers of strings with nearly opposite directions reconnecting.

The nonrelativistic limit also makes it easy to see the connection to magnetohydrodynamics. The vector $\flux\B{w}$ is just the magnetic field $\B{B}$, and from the equation of state \eqref{mhdMu} $\mu=\flux$. So from \eqref{faraday} the electric field vector is equal to
\begin{align}
\B{E}=\B{B}\times \B{v}+\B{w}\times\B{\nu}+\tilde{\B{G}}.
\end{align}
Bringing $\mu=\flux$ inside the curls in \eqref{gtilde},
\begin{align}
\B{E}=\B{B}\times \B{v}+\frac{\coeff_T}{T}(\nabla\times\B{B})_\perp+\frac{\coeff_T}{T^2}\B{B}\times\nabla T+\frac{\coeff_L}{T}  (\nabla\times\B{B})_w.\label{efield}
\end{align}
The first term is also in ideal magnetohydrodynamics and is due to the Lorentz boost out of the rest frame of the fluid. At low frequencies the displacement current can be neglected and Ohm's law can be written
\begin{align}
\B{E}=\sigma^{-1}\B{J}=\sigma^{-1}\nabla\times\B{B}.
\end{align}
So the coefficients $\coeff$ can be related to the electrical conductivity $\sigma$,
\begin{align}
\coeff = \frac{T}{\sigma}.
\end{align}
This is somewhat different from ordinary resistive magnetohydrodynamics due to the possibility of anisotropic conductivity, but also due to the presence of the temperature gradient term. In the string Eckart frame this term would vanish, but that would also restrict $E$ to be parallel to $B$ in the rest frame.

The origin of the difference can be seen by comparing our introduction of dissipative terms in this paper to the standard introduction of Ohm's law discussed in Sec. \ref{mhd}. In standard MHD the energy-momentum tensor is assumed to be separated into distinct fluid and electromagnetic parts $T_\text{m}+T_\text{EM}$  even out of equilibrium. The entropy is taken to only be a function of the fluid quantities, not the electromagnetic part. This makes sense in equilibrium since dependence on $ \flux $ and the electromagnetic energy density cancel
\begin{align}
ds &= \frac{1}{T}d\rho_{\text{total}} - \frac{\mu}{T}d\flux - \frac{m^a}{T}dn_a\non
&=\frac{1}{T}d\rho_m  - \frac{m^a}{T}dn_a.
\end{align}
But the string fluid approach taken in this paper has entropy be a function of electromagnetic sector out of equilibrium, leading to the presence of a term in the entropy current representing the diffusion of field lines \eqref{scurrent}. This diffusion term in the entropy current is ultimately responsible for the presence of the temperature gradient term in the electric field \eqref{efield}.

\subsection{Stationary solutions}\label{secEquil}
If a dissipative string fluid reaches a state of maximum entropy, the requirement that no further entropy be produced leads to stricter restrictions than are found in the perfect string fluid. This is a direct analogy to the stationary solutions of ordinary relativistic fluids which have among other things been taken to model rotating stars \cite{LindblomRotating}.

Clearly for the entropy to be conserved all of the dissipative terms leading to entropy production in \eqref{deriv2} must vanish. For the components of the viscous stress $\pi^{\mu\nu}$ to vanish, the shear and expansion $\nabla_{(\mu} u_{\nu)}$ must also vanish. In particular,
\begin{align}
\nabla_\mu u^\mu &= 0\\
w^{\mu}w^\nu\nabla_\mu u_\nu &= 0.\label{expansion}
\end{align}
So the conservation of entropy \eqref{ds} and the vanishing of expansion implies $s$ is constant in the flow direction
\begin{align}
u^\mu\nabla_\mu s = 0.\nonumber
\end{align}
If there are any conserved currents $n_a$ besides the entropy clearly these must also be constant in the $u$ direction by the same reasoning. Furthermore, using the vanishing of shear \eqref{expansion} in the expression for the divergence of $F$:
\begin{align}
0=w_\mu\nabla_\nu F^{\mu\nu}=-\nabla_\nu \flux u^\nu.\label{fluxU}
\end{align}
So $\flux$ is also constant in the flow direction, and thus all thermodynamic variables must be. 

To proceed we will make use of a general relation for perfect string fluids. From the contracted conservation of $T$,
\[ w_\mu\nabla_\nu T^{\mu\nu}=0, \]
it can be shown that the `dual currents' $m^a w$ satisfy the relation
\begin{align}
s\nabla_\mu \flux T w^\mu + n_a\nabla_\mu \flux m^a w^\mu = 0.\nonumber
\end{align}
Incidentally, this is a fluid generalization of the dual current which appears in Carter's work on single strings \cite{CarterDuality}. For simplicity the following demonstration will consider the case where the entropy is the only current so that
\begin{align}
\nabla_\mu \flux T w^\mu  = 0.\label{tW}
\end{align}

Beginning with the conservation of $T$, and making use of the relation above and the conservation of $\rho+p$ in the $u$ direction:
\[0=\nabla_\mu T^{\mu\nu}= (\rho+p)u^\mu\nabla_\mu u^\nu -\flux T w^\mu \nabla_\mu \frac{\mu}{T} w^\nu  - \nabla^\nu p.\]
The requirement that the diffusion vector $\nu$ vanishes implies
\begin{align}
\perp^{\lambda\nu}(w^\mu\nabla_\mu w_\nu+\nabla_\nu \,\text{ln}\,\frac{\mu}{T})=0,\label{nuStationary}
\end{align}
so then the conservation of $T$ can be simplified further to
\begin{align}0=(\rho+p)u^\mu\nabla_\mu u^\nu +\flux T \nabla^\nu \frac{\mu}{T} -\nabla^\nu p.\nonumber
\end{align}
 Making use of the thermodynamic identity \eqref{thermIden}, this implies
 \begin{align}
 u^\mu\nabla_\mu u^\nu = \nabla^\nu \,\text{ln}\,T, \label{logT}
 \end{align}
which together with the vanishing of the shear of $u$ leads to the conclusion
\begin{align}
\nabla_{(\mu}\beta_{\nu)}=\nabla_{(\mu}\frac{1}{T}u_{\nu)}=0.
\end{align}
So $\beta$ is a Killing vector in equilibrium, a fact also true for ordinary fluids.

At this point, note that the orthogonal projection of $\nabla_\nu T^{\mu\nu}=0$
leads to
\begin{align}
\perp^{\lambda\mu}(u^\nu\nabla_\nu u_\mu - w^\nu\nabla_\nu w_\mu - \nabla_\nu \,\text{ln}\,\mu)= 0.\label{curvStationary}
\end{align}
The first two terms have a natural interpretation as the extrinsic curvature vector $K$,
\begin{align}
K^\lambda &\equiv h^{\rho}_{\,\,\,\sigma}\nabla_\rho h^{\sigma\lambda}=\perp^{\lambda\rho}(u^\sigma\nabla_\sigma u_\rho- w^\sigma\nabla_\sigma w_\rho).\nonumber
\end{align}
So in the stationary solutions, curvature in the macroscopic worldsheets is balanced by changes in $\mu$. This relation \eqref{curvStationary} was noticed already in \cite{blackfold} through a different line of reasoning. In our approach the similar relation \eqref{nuStationary} relating the curvature of the field lines to changes in $\mu/T$ is more quickly seen.

At equilibrium there is a Killing vector $\beta$ in the direction of the velocity $u$. It will turn out there is also a preferred vector in the $w$ direction. Using the conservation of $F$ and \eqref{fluxU}\eqref{tW},
\begin{align}
0=\nabla_\mu F^{\mu\nu}&=\flux T w^\mu \nabla_\mu \frac{1}{T}u_{\nu} - \flux u^\mu \nabla_\mu w^\nu\non
&=\flux(u^\nu \nabla_\mu w_\nu - u^\nu \nabla_\nu w_\mu),\nonumber
\end{align}
where the Killing vector property was used in the second line. Therefore it is true that
\[ u^\mu\nabla_{[\mu}\frac{\mu}{T}w_{\nu]}=0, \]
and using the vanishing of $\nu$ and $G$ (which depend on the other components),
\begin{align}
\nabla_{[\mu}\alpha_{\nu]} =\nabla_{[\mu}\frac{\mu}{T}w_{\nu]}=0.
\end{align}

So $\alpha$ and $\beta$, which were introduced earlier as derivatives of the entropy, form a natural coordinate system for the stationary fluid. The fact that their commutator vanishes can be easily proven from the conservation of $F$ as above. Note that this is distinct from the analysis of a preferred spacelike vector appearing in \cite{blackfold}. There the assumption that all thermodynamic quantities are constant along the field lines $w$ was effectively made, restricting the generality of the stationary solutions.

Finally we note that as for the case of an ordinary fluid, $m^a/T$ for each current is constant throughout the fluid. This follows easily from the vanishing of the dissipative part of $n_a$ in the Landau-Lifshitz frame \cite{LL}.

\subsection{Second-order theory}\label{sec2nd}

The theory we have been discussing is essentially an extension of the `first-order' relativistic fluids of Eckart \cite{Eckart} and Landau-Lifshitz \cite{LL}. It is well known that these theories suffer certain difficulties. Hiscock and Lindblom have shown that the equilibrium states are unstable on short time scales under certain perturbations \cite{HLInstability}. Another difficulty of first-order theories which is easily seen to be present in the current theory as well is the appearance of parabolic equations. For instance, the equation for longitudinal heat flow is given by \eqref{Q}
\[ \Q = \kappa_L w^\mu (\nabla_\mu T - T u^\nu \nabla_\nu u_\mu). \]
For a system of straight strings at rest with no orthogonal gradients, this leads to the one-dimensional heat equation
\[ \dot{T}=\frac{\kappa_L}{C}\partial_w^2 T, \]
where $C$ is the heat capacity at constant flux
\begin{align}
C \equiv \der {\rho}{T}.
\end{align}
So a small perturbation in $T$ will instantly be felt across the entire string.

The resolution to both problems for ordinary fluids \cite{israelStewart}\cite{HL1983} is by including second-order terms in expansion of the nonequilibrium entropy current $s^\mu$ \eqref{scurrent}. For instance, an additional term $-\frac{1}{2}k u^\mu \Q^2 $ for some positive coefficient $k$ will lead to an extra term $-\kappa_L T^2 k \dot{Q}_L$ in the expression for heat conduction above \eqref{Q}. This will in turn modify the heat equation to
\begin{align}
kCT^2\ddot{T}+\frac{C}{\kappa_L}\dot{T}=\partial_w^2 T,\nonumber
\end{align}
which is now hyperbolic, with the speed of second sound
\begin{align}
c_s^2 \equiv \frac{1}{kCT^2 }.\label{secondsound}
\end{align}

As a practical matter however, there are many more possible independent second-order terms in the string fluid than in the ordinary Israel-Stewart theory. This is both due to the breaking of rotational symmetry into transverse and longitudinal directions, and also due to the presence of an extra direction in equilibrium. For instance there may be all the possible terms,
\[ g_{\rho\sigma}\nu^\rho \pi_L^\sigma u^\nu, g_{\rho\sigma}\nu^\rho \pi_L^\sigma w^\nu, \tS_{\rho\sigma}\nu^\rho \pi_L^\sigma u^\nu,\dots\]
and so on ---each with an independent parameter.

Even so there are some principles which can restrict the number of independent terms. For one it should be required that the theory be invariant under changes of frame. The full entropy current $s$ is a function of the tensors $T$ and $F$, but we have expanded it about a certain arbitrary equilibrium state $T_0,F_0$. Expanding about a different equilibrium state should lead to the same result to the order of the highest term kept in the expansion.

Following the same approach as Israel-Stewart \cite{israelStewart}, the entropy current \eqref{scurrent} is given a second-order correction $S$,
\begin{align}
s^\mu = {p}\beta^\mu + \beta_\nu T^{\mu\nu} - \alpha_\nu F^{\mu\nu} - \frac{m^a}{T} n_{a}^\mu + S^\mu.
\end{align}
The principle of frame invariance is then that $ ds^\mu = 0 $ under changes of $u$ and $w$.

The thermodynamic relation \eqref{s0} may be Legendre transformed to
\begin{align}
d(p\beta^\mu)=F_0^{\mu\nu}d\alpha_\nu - T_0^{\mu\nu}d\beta_\nu + n_0^\mu d(\frac{m^a}{T}).
\end{align}
So the change in $s$ under changes of $\alpha,\beta$ is
\[ ds^\mu = (T-T_0)^{\mu\nu}d\beta_\nu - (F-F_0)^{\mu\nu}d\alpha_\nu + dS^{\mu}, \]
and by frame invariance the change in $S$ must be,
\begin{align}
dS^{\mu} = \frac{\mu}{T}(F-F_0)^{\mu\nu}dw_\nu - \frac{1}{T}(T-T_0)^{\mu\nu}du_\nu.\nonumber
\end{align}
Using the full decomposition of the tensors in Sec. \ref{sectTDecomp}, this is
\begin{align}
dS^{\mu} = &\frac{\mu}{T}(-u^\mu \lam^\nu+w^{\mu}\nu^\nu+G^{\mu\nu})dw_\nu \non
&- \frac{1}{T}(u^\mu q_T^\nu-w^\mu \pi_L^\nu + \pi_T^{\mu\nu} +\Pi_T \perp^{\mu\nu})du_\nu \non
&- \frac{1}{T}(\mu\lam^\mu + \Q u^\mu - \pi_L^\mu - \Pi_L w^\mu)w^\nu du_\nu.\label{dS}
\end{align}
So $S$ may include arbitrary terms which are invariant to second order under changes of frame, but it must also include terms so as to produce the change above.

Clearly it is important to know how the various quantities change with the frame. The changes $du,dw$ to nearby equilibrium states are on the order of the field quantities themselves, as can be seen for instance in the change to the Eckart velocity in \eqref{eckart}. The thermodynamic quantities $\rho,\flux,n_a$ defined through \eqref{rho} are all invariant to first order, and thus so must be any thermodynamic quantity. Likewise $G^{\mu\nu},\Pi_L,\Pi_T,\pi_T^{\mu\nu}$ are all invariant to first order, but the remaining dissipative fields are not:
\begin{align}
d\nu^\mu &= -\flux du_\perp ^\mu\non
dq_T^\mu &= -(\rho+p)du_\perp ^\mu\non
d\lam^\mu &= -\flux dw_\perp ^\mu\non
d\pi_L^\mu &= -(\tau+p)dw_\perp ^\mu\non
d\Q &= +(\rho-\tau)w^\nu du_\nu,\label{variant}
\end{align}
where the subscript $\perp$ indicates the change is projected orthogonal to $u,w$.

Even though these are not invariant, they can form the invariant combinations
\begin{align}
q_T^\mu-\frac{\rho+p}{\flux}\nu^\mu\\
\pi_L^\mu -\mu\lam^\mu.\label{invariant}
\end{align}
This is a very modest step in reducing the complexity of the second-order theory in that the five quantities in \eqref{variant} may only appear with arbitrary parameters in the two combinations above. Note that the first combination, the invariant heat, was implicitly already used in \eqref{eckartHeat} to relate $\nu$ to heat conduction.

The change in $S$ \eqref{dS} can only be produced by the noninvariant terms \eqref{variant}, and we will denote this noninvariant piece $S_0$. There is some ambiguity in how to split this from the invariant part of $S$, but we will make a choice so that $S_0$ vanishes in the Landau-Lifshitz frame. It can then be explicitly calculated:
\begin{align}
S_0^\mu = &\frac{1}{T}(\frac{1}{2}u^\mu\q_T^\nu  - w^\mu\pi_L^\nu +\pi_T^{\mu\nu}+\mu w^\mu\lam^\nu)\frac{q_{T\,\nu}}{\rho+p}+\frac{\mu}{ T}(\frac{1}{2}u^\mu \lam^\nu-w^\mu \nu^\nu)\frac{\lam_\nu}{\flux}\non
&-\frac{1}{T}(\frac{1}{2}\Q u^\mu -\Pi_L w^\mu -\pi_L^\mu+\mu\lam^\mu)\frac{\Q}{\rho-\tau}.
\end{align}
In the absence of any particle currents the longitudinal heat $\Q$ transforms differently from the other quantities \eqref{variant}. So its only appearance in the second-order theory is in the terms of $S_0$ above, with no new parameters. 

Thus the coefficient $k$ of the $Q_L^2$ term which leads to the speed of second sound \eqref{secondsound} is
\begin{align}
k = \frac{T^{-1}}{\rho-\tau}=\frac{1}{sT^2},\nonumber
\end{align}
where the second equality uses the fact that there are no particle currents in the equation of state. So the speed of second sound is
\begin{align}
c_s^2= \frac{s}{C} &= \frac{s}{T}\der{T}{s}.
\end{align}
In a pressureless perfect string fluid this is just the expression for the ordinary longitudinal speed of sound (see for instance \cite{carterWave}).

 In particular, recalling the wiggly string fluid equation of state \eqref{wigglyState}
\[ \rho = \sqrt{(\mu_0 \flux)^2 + (T_H s)^2}, \]
the speed of second sound is
\begin{align}
c_s =\sqrt{\frac{\tau}{\rho}} =\sqrt{1-\left(\frac{T}{T_H}\right)^2}.
\end{align}
This is again just equal to the ordinary speed of perturbations on the string, expressed in terms of the tension and mass density. And the second equality makes it clear that the speed of second sound is causal and vanishes as the temperature approaches the Hagedorn temperature. Of course for many reasons the wiggly string fluid equation of state should be understood as a toy model, but this reasonable result is at the very least a consistency check on the second-order theory.

\section*{Acknowledgments}

I would like to thank Vitaly Vanchurin for interesting discussions, and for previously helping notice the connection to electromagnetism that is expanded upon here.

\end{document}